# Orbital-selective two-gap superconductivity in kagome metal $CsV_3Sb_5$

Run Lv[1,2], Wenqian Tu[1,2], Dingfu Shao[1], Yuping Sun[3,1,4], and Wenjian Lu[1,*]

[1]Key Laboratory of Materials Physics, Institute of Solid State Physics, HFIPS, Chinese Academy of Sciences, Hefei 230031, China
[2]University of Science and Technology of China, Hefei 230026, China
[3]High Magnetic Field Laboratory, HFIPS, Chinese Academy of Sciences, Hefei 230031, China
[4]Collaborative Innovation Center of Microstructures, Nanjing University, Nanjing 210093, China

Recent experiments have revealed anisotropic multi-gap superconductivity in the kagome metal $CsV_3Sb_5$. However, the impact of multi-orbital character and electron-phonon coupling (EPC) on the multiple superconducting gaps remains not fully understood. In this work, we theoretically investigate the multi-orbital electronic structure and superconducting gap properties of $CsV_3Sb_5$ by combining first-principles calculations with superconducting density functional theory (SCDFT). Our results demonstrate that orbital-selective pairing drives the observed two-gap superconductivity in $CsV_3Sb_5$. Specifically, the two distinct gaps exhibit strong orbital dependence: a large, highly anisotropic gap (average magnitude ~0.64 meV) primarily originates from V-$3d$ orbitals, while a small, isotropic gap (~0.25 meV) is associated with the Sb-$5p_z$ orbital. The V-$3d$ orbitals exhibit strong coupling to the in-plane V-V bond-stretching and out-of-plane V-Sb bending phonon modes, whereas the Sb-$5p_z$ orbitals show weak coupling to the out-of-plane vibrations of both Cs and the apical Sb atoms. These findings provide fundamental insights into the orbital-selective superconductivity and EPC mechanisms in kagome $CsV_3Sb_5$.

[*]Corresponding author: wjlu@issp.ac.cn

## I. INTRODUCTION

The recently discovered kagome metal $AV_3Sb_5$ (A=K, Rb, Cs) family exhibits rich quantum phenomena, including charge density wave (CDW) [1-11], pressure-tunable superconductivity [12-26], pair density wave [27,28], anomalous Hall effect [29], and time-reversal symmetry breaking [30]. Among the $AV_3Sb_5$ family, $CsV_3Sb_5$ has emerged as a primary research platform for investigating the unconventional superconductivity owing to its relatively high superconducting transition temperature ($T_C \sim 2.5$ K) [13], and the coexistence of multiple quantum states [13,21,29,31-35]. Across various experimental techniques, the pairing symmetry has been consistently pointed toward an $s$-wave spin-singlet state in $CsV_3Sb_5$ [16,17,19,22,25,36]. Moreover, the importance of electron-phonon coupling (EPC) in superconducting pairing is supported by theoretical calculations [37-39], and is further corroborated by the experimental observations of EPC-induced kinks in the electronic band structure [40,41].

Evidence of two-gap superconductivity in $CsV_3Sb_5$ has been reported in different experiments, such as transport experiments [18,42], scanning tunneling spectroscopy (STS) [19], and point-contact spectroscopy (PCS) experiments under pressure [20,23,26]. Meanwhile, the V-shaped superconducting gaps [5,19] and finite residual density of states (DOS) at zero energy [43,44] are consistently observed. Magnetic penetration depth [25] and electronic specific heat [45] experiments further revealed the coexistence of an isotropic gap and an anisotropic gap. Despite recent progress, the microscopic origin of these complex multi-gap features in $CsV_3Sb_5$ is not yet fully understood.

In this work, we investigate the superconducting gap properties of $CsV_3Sb_5$ using first-principles calculations combined with superconductivity density functional theory (SCDFT), a parameter-free approach that avoids the empirical Coulomb pseudopotential $\mu^*$. Our results reveal that EPC in $CsV_3Sb_5$ is highly orbital-selective: the V-3$d$ orbitals strongly couple to the in-plane V-V bond-stretching and out-of-plane V-Sb bond-bending phonon modes, whereas the Sb-5$p_z$ orbitals weakly couple to the out-of-plane vibrations of both Cs and the apical Sb atoms. This orbital selectivity dictates the superconducting pairing: V-3$d$ orbitals generate a large and anisotropic gap with an average value of 0.64 meV, while the Sb-5$p_z$ orbitals yield a small and isotropic gap around 0.25 meV. Moreover, the EPC-induced band renormalization manifests as distinctive kinks at approximately -13 meV and -30 meV in the electronic dispersion, consistent with angle-resolved photoemission spectroscopy (ARPES) observations. Our results establish that orbital selectivity drives the two-gap superconductivity in $CsV_3Sb_5$, thereby contributing to elucidating the mechanism of unconventional superconducting pairing.

## II. METHODS

The first-principles calculations based on the density functional theory (DFT) were performed using the Quantum Espresso package [46,47]. The ultrasoft pseudopotentials were used to describe the interaction between electrons and ionic cores [48]. The exchange-correlation interaction was described by the generalized gradient approximation (GGA) and parameterized by the Perdew-Burke-Ernzerhof (PBE) functional [49]. The cutoffs of kinetic energy and charge density were set to 50 and 500 Ry, respectively. The van der Waals (vdW) interaction was included by the nonlocal rVV10 method [50,51]. The structure was fully relaxed until the Hellmann-Feynman force acting on each atom was less than $10^{-4}$ Ry/Å, and the convergence criterion for self-consistent calculations was set to be $10^{-6}$ Ry. The Brillouin zone (BZ) was sampled with a 20×20×12 $k$-points grid and a 5×5×3 $q$-points grid for electronic and phonon calculations, respectively. The EPC renormalized band structure and electron self-energy calculations were carried out using the Electron-Phonon Wannier (EPW) code [52]. The Matsubara frequency cutoff was set as 200 meV. The momentum-resolved superconducting gaps were calculated by the Superconducting-Toolkit (SCTK) code [53-55] based on the SCDFT [56]. As required by this method, the electron-phonon calculations were performed using a 5×5×3 shifted $q$-points grid, yielding 18 symmetry-inequivalent $q$-points. The tetrahedron method was used for all BZ integrations.

## III. RESULTS AND DISCUSSION

$CsV_3Sb_5$ exhibits the kagome lattice at ambient conditions, with a hexagonal space group of $P6/mmm$ (No. 191). Its structure is formed by V-Sb slabs stacked along the $c$-axis, separated by intercalated Cs atoms, as illustrated in Fig. 1(a). The V-Sb slabs are composed of in-plane V and Sb atoms, which form the kagome plane, and apical Sb atoms. At low temperatures, $CsV_3Sb_5$ exhibits the coexistence of superconductivity and CDW. Upon increasing pressure, the CDW order is monotonically suppressed and disappears at 2 GPa [8,14,15,24,35]. The presence of CDW-related imaginary (negative) frequencies of the phonon spectrum for the ideal kagome structure at ambient/low pressures precludes the definition of EPC strength. Thus, our electronic band, phonon, EPC, and SCDFT calculations were performed at 3 GPa, at which the CDW is fully suppressed.

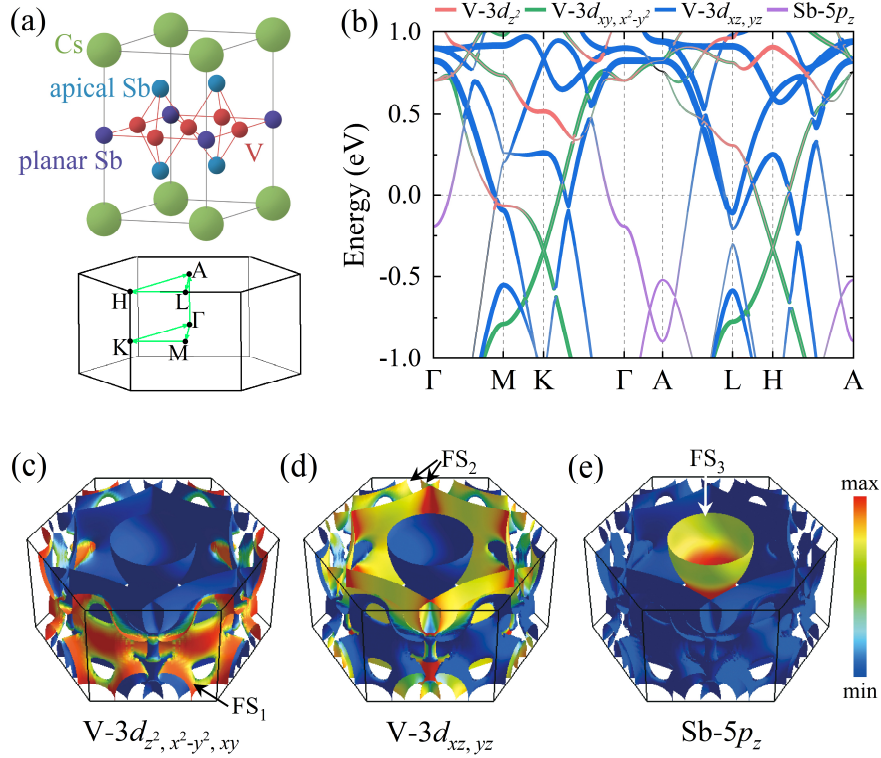

FIG. 1. (a) Crystal structure and Brillouin zone of $CsV_3Sb_5$. (b) Orbital-resolved band structure, and (c)-(e) Fermi surfaces of $CsV_3Sb_5$ colored with orbital-weight distribution.

Figure 1(b) presents the electronic band structure of $CsV_3Sb_5$. Two bands cross the Fermi level, primarily derived from V-$3d$ and Sb-$5p$ orbitals. Around the $M$ point, the V-$3d_{z^2}$ and $3d_{x^2-y^2,xy}$ orbitals hybridize, giving rise to a van Hove singularity. The momentum distributions of these orbitals are presented in Figs. 1(c)-(e), with additional atom- and orbital-resolved band structures and FS provided in Fig. S1 of the Supplemental Material. As shown in Fig. 1(c), the V-$3d_{z^2,x^2-y^2,xy}$ orbitals are predominantly localized on the outer Fermi surface sheet, denoted as $FS_1$. Figure 1(d) reveals that the hexagonal $FS_2$ mainly arises from V-$3d_{xz,yz}$ orbitals, except in the region around the *M* point. The $5p_{x,y}$ orbitals of the apical Sb atoms exhibit very weak hybridization with the V-$3d_{xz,yz}$ orbitals (see Fig. S1(i)). The electron pocket around the $\Gamma$ and *A* points is primarily formed by $5p_z$ orbitals from both planar and apical Sb atoms, constituting a cylindrical FS$_3$ enclosing the $\Gamma$-*A* direction. Additionally, FS$_3$ receives a minor contribution from the V-$3d_{xz,yz}$ orbitals near the $k_z$=0.5 plane (see Fig. S1(d)). These FS sheets show slight variations in morphology and orbital composition across different $k_z$ values, indicating a quasi-two-dimensional character.

Multiple experimental studies based on transport measurements [16,18,25,36], STM [5,19], and nuclear magnetic resonance/nuclear quadrupole resonance (NMR/NQR) [17] consistently support *s*-wave singlet pairing in CsV$_3$Sb$_5$. This consensus suggests conventional superconductivity, which is expected to be mediated by EPC. Indeed, subsequent first-principles calculations incorporating EPC provide strong support for phonon-mediated superconductivity [37-39]. To study the superconducting gap of CsV$_3$Sb$_5$, we start by calculating the phonon dispersion and EPC properties. Figure 2(a) shows the phonon spectrum weighted by the magnitude of phonon linewidth $\gamma_{qv}$, which provides information on EPC. The phonon linewidth is calculated by

$$\gamma_{qv} = 2\pi\omega_{qv} \sum_{ij} \int \frac{d^3k}{\Omega_{BZ}} |g_{qv}(k,i,j)|^2 \times \delta(\varepsilon_{q,i} - \varepsilon_F)\delta(\varepsilon_{k+q,i} - \varepsilon_F), \qquad (1)$$

where $g_{qv}$ is the EPC matrix and can be calculated using the formula

$$g_{qv}(k,i,j) = \left(\frac{\hbar}{2M\omega_{qv}}\right)^{1/2} \langle \psi_{i,k} | \frac{dV_{SCF}}{d\hat{u}_{qv}} \hat{\xi}_{qv} | \psi_{j,k+q} \rangle. \qquad (2)$$

Here, $\psi_{i,k}$ is the wave function, $\frac{dV_{scf}}{d\hat{u}_{qv}}$ is the change of the Kohn-Sham potential caused by atomic displacement $\hat{u}_{qv}$, and $\hat{\xi}_{qv}$ is the phonon eigenvector. The width of the lines in blue color in Fig. 2(a) represents the magnitude of $\gamma_{qv}$. Within the low-frequency region ($\omega$<15 meV), the large phonon linewidth primarily originates from the softening of phonon modes around the *M* and *L* points. To illustrate the orbital-dependent EPC, schematic diagrams of atomic vibrations corresponding to major soft phonon modes are presented in Figs. 2(d)-(f). For further comparison, the vibration-resolved phonon spectra are provided in Fig. S2 of the Supplemental Material. In the high-frequency regime ($\omega$>15 meV), two soft modes emerge near the *M* and *L* points, both associated with the out-of-plane V-Sb bond-bending phonon mode (Fig. 2(d)). In the low-frequency regime ($\omega$<15 meV), a prominent soft phonon mode around the *M* point is linked to the in-plane V-V bond-stretching vibrations within the kagome plane, accompanied by the out-of-plane vibration of apical Sb atoms (Fig. 2(e)). Additionally, two soft phonon modes near the *L* point display similar variations of V and Sb atoms to the phonon mode around the *M* point, but also incorporate a shearing motion of Cs atoms (Fig. 2(f)). The out-of-plane vibrations of both Cs and apical Sb atoms make a non-negligible contribution to the low-frequency soft modes around the *L* point (see Fig. S2(b) of the Supplemental Material).

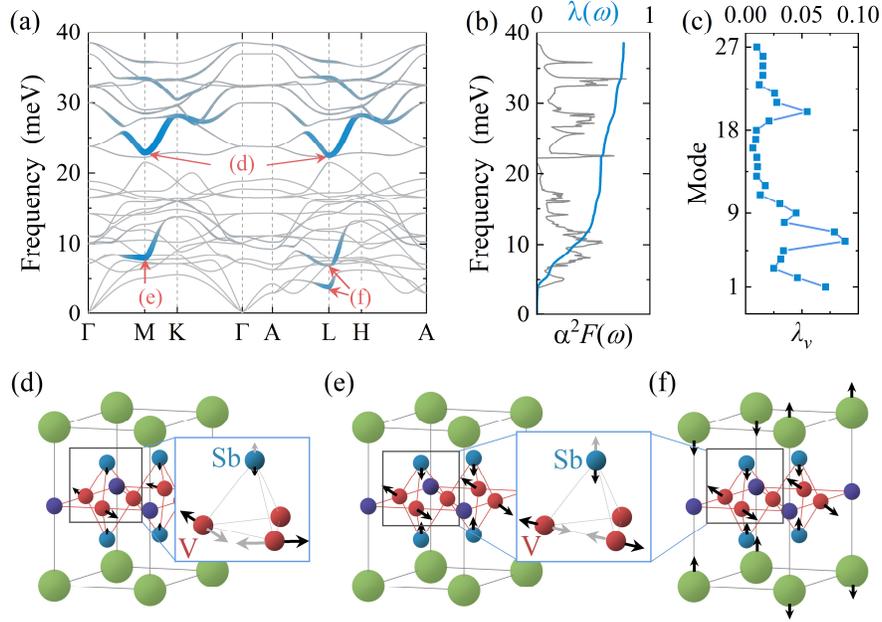

FIG. 2. (a) Phonon linewidth $\gamma_{qv}$ (blue color) decorated phonon spectra. The atomic vibration schematics of soft phonon modes marked by red arrows are shown in (d)-(f). (b) Eliashberg spectral function $\alpha^2F(\omega)$ and integrated EPC strength $\lambda(\omega)$. (c) EPC strength of each phonon mode.

To evaluate the strength of EPC, we calculated the Eliashberg spectral function $\alpha^2F(\omega)$ and EPC spectrum $\lambda(\omega)$ by

$$\alpha^2 F(\omega) = \frac{1}{2\pi N(E_F)} \sum_{qv} \delta(\omega - \omega_{qv}) \frac{\gamma_{qv}}{\hbar \omega_{qv}} \quad (3)$$

and

$$\lambda(\omega) = 2 \int \frac{\alpha^2 F(\omega)}{\omega} d\omega. \quad (4)$$

Figure 2(b) presents the corresponding calculation results. The total integrated EPC strength $\lambda$ is 0.77. Contributions from high-frequency phonon modes ($\omega$>15 meV) and low-frequency modes ($\omega$<15 meV) are 0.25 and 0.52, respectively. In the high-frequency regime, the EPC strength primarily stems from the coupling between V-$3d_{xz,yz}$ orbitals and the out-of-plane V-Sb bond-bending mode. Although these modes exhibit significant softening and large linewidths $\gamma_{qv}$, their high frequencies result in only a moderate contribution to the total EPC strength. In the low-frequency regime, the soft modes are dominated by the in-plane V-V bond-stretching phonon mode, with minimal contributions from out-of-plane vibrations of both Cs and apical Sb atoms around the $L$ point (see Fig. S2(b)). Accordingly, the EPC strength originates mainly from the coupling between V-$3d_{z^2,x^2-y^2,xy}$ orbitals and V-V bond-stretching phonon mode. The Sb-$5p_z$ orbitals couple to the out-of-plane vibrations of both Cs and apical Sb atoms, providing a relatively small yet non-negligible contribution to the EPC strength, consistent with a previous report [57]. The remaining phonon modes contribute negligibly to the total EPC strength, as illustrated in Fig. 2(c). Furthermore, based on the Allen-Dynes-modified McMillan formula

$$T_c = \frac{\omega_{log}}{1.2}\exp(-\frac{1.04(1+\lambda)}{\lambda-\mu^*-0.62\lambda\mu^*}), \quad (5)$$

we estimated $T_C$ to be 5.8 K at 3 GPa, which agrees with the previously experimentally obtained values of 4.2-6.6 K [8,15,26,43].

To investigate the impact of EPC on electron properties in CsV$_3$Sb$_5$, we calculated the EPC-renormalized electronic band structure, which can be corroborated by the ARPES measurements. In ARPES experiments, the effects of many-body interactions, such as EPC, can be observed in the spectral function $A_{nk}(\omega)$, expressed as

$$A_{nk}(\omega) = \frac{1}{\pi}\frac{|Im\,\Sigma_{nk}(\omega)|}{|\omega-\varepsilon_{nk}-Re\,\Sigma_{nk}(\omega)|^2+|Im\,\Sigma_{nk}(\omega)|^2}. \quad (6)$$

Here $\varepsilon_{nk}$ stands for the non-interacting bare band, or the Kohn-Sham eigenvalue in DFT calculations [41,58]. $\Sigma_{nk}(\omega)$ represents single-particle self-energy and contains the information of the interaction between electron and phonon. One can calculate the electron self-energy caused by EPC and study its modifications of band structures. The self-energy can be calculated as

$$\Sigma_{nn'k}(\omega) = \frac{1}{\hbar}\Sigma_{mv}\int\frac{d\mathbf{q}}{\Omega_{BZ}}g_{mnv}^*(\mathbf{k},\mathbf{q})g_{mn'v}(\mathbf{k},\mathbf{q})$$
$$\times[\frac{1-f_{mk+q}+n_{qv}}{\omega-\varepsilon_{mk+q}/\hbar-\omega_{qv}+i\eta}+\frac{f_{mk+q}+n_{qv}}{\omega-\varepsilon_{mk+q}/\hbar+\omega_{qv}+i\eta}], \quad (7)$$

where $n(\omega_{qv})$ and $f(\varepsilon_{mk+q})$ denote the Bose and Fermi occupation factors [56], respectively. Comparing Eqs. (7) and (1), one can find that the electronic self-energy $\Sigma_{nk}$ is also proportional to the $\Sigma g_{qv}^2$, and thus the influence of EPC could be observed in the spectral function $A_{nk}(\omega)$.

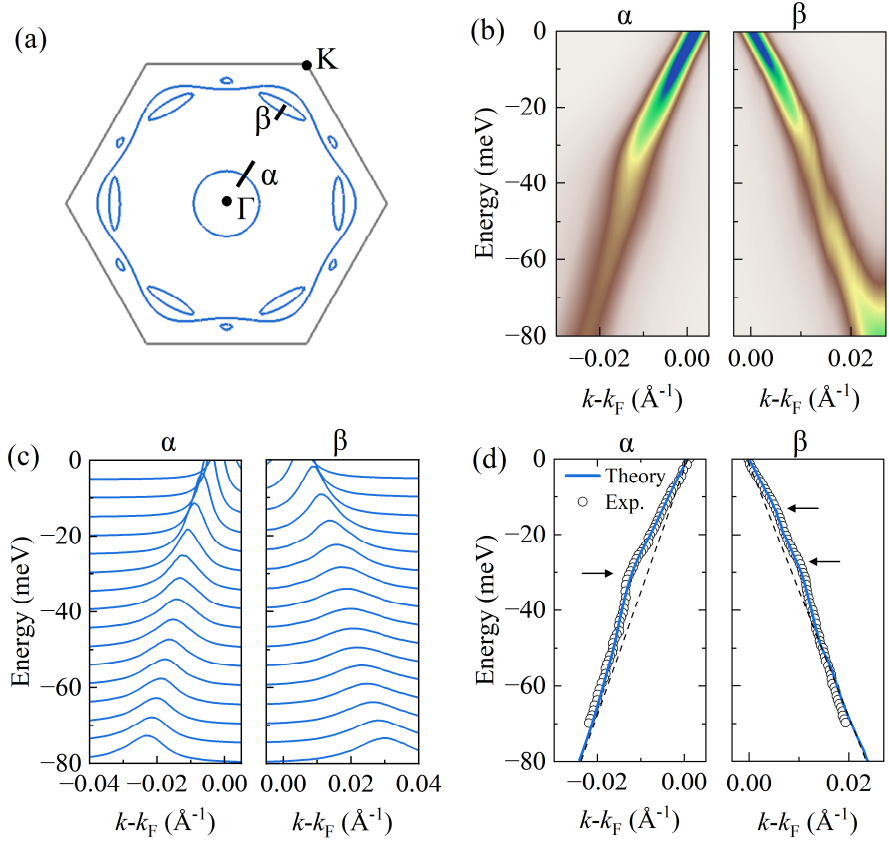

FIG. 3 (a) Section of Fermi surface at $k_z$=0 plane. The black lines show the momentum location of the cuts. (b) Spectral function $A(\mathbf{k}, \omega)$ of the EPC renormalized band structure. (c) Extracted momentum dispersion curves and (d) dispersions for the α and β bands. Positions of kinks are indicated by black arrows. Experimental data are extracted from Ref. [41].

In the previous ARPES measurements, clear kinks were observed in the electron spectral function and attributed to the EPC effect [41]. To provide a theoretical understanding, we calculate the EPC-renormalized band structure near the Fermi level. The results are presented in Fig. 3. For comparison with experimental results, we selected the same two cuts along the $\Gamma$-$K$ direction of the Fermi surface, as indicated by the black lines in Fig. 3(a). Compared to the Kohn-Sham band shown in Fig. 1(b), the dispersion of bands is not smooth anymore. We extracted the momentum dispersion curves (MDC) from the spectral function, using their peak positions to determine the renormalized band dispersion. The dashed lines in Fig. 3(d) indicate the Kohn-Sham band structure. After considering EPC, a kink appears near -31 meV in the α band, and two kinks appear around -13 meV and -30 meV in the β band, consistent with the observations in ARPES experiments [41]. The energies of the kinks align closely with the frequencies of soft phonon modes with strong EPC, confirming EPC as the microscopic origin of the observed band renormalizations.

To gain further insight into the superconducting properties of $CsV_3Sb_5$, we employed the parameter-free SCDFT to calculate the momentum-resolved superconducting gap $\Delta_{nk}$. Within the SCTK package implementation, the superconducting gap function $\Delta_{nk}$ at temperature $T$ is described by [53]

$$\Delta_{nk} = -\frac{1}{2}\sum_{n'k'} \frac{K_{nkn'k'}(\xi_{nk},\xi_{n'k'})}{1+Z_{nk}(\xi_{nk})} \times \frac{\Delta_{n'k'}}{\sqrt{\xi^2_{n'k'}+\Delta^2_{nk}}} \tanh \frac{\sqrt{\xi^2_{n'k'}+\Delta^2_{nk}}}{2T}, \qquad (8)$$

where $\xi_{nk}$ is the Kohn-Sham eigenvalue of the $n$'th band near the Fermi energy at the wave vector $k$. $K_{nkn'k'}(\xi_{nk},\xi_{n'k'})$ is the integration kernel, indicating the superconducting-pair creating and breaking interactions, and is defined by

$$K_{nkn'k'}(\xi,\xi') = K^{ep}_{nkn'k'}(\xi,\xi') + K^{ee}_{nkn'k'}(\xi,\xi') + K^{sf}_{nkn'k'}(\xi,\xi'), \qquad (9)$$

consisting of the electron-phonon, the Coulomb repulsion, and the spin fluctuation kernel terms. The superconducting gap function (Eq. (8)) is solved numerically at each temperature, and therefore, $T_C$ can be obtained when the gap vanishes. This approach directly incorporates the Coulomb repulsion $K^{ee}$, and does not rely on an empirical value of Coulomb pseudopotential $\mu^*$.

Figure 4(a) shows the calculated momentum-resolved superconducting gap $\Delta_{nk}$. The gaps on the $FS_1$ and $FS_2$ are highly anisotropic, while the gap shows an isotropic distribution on the cylindrical $FS_3$ around the $\Gamma$ point. The $FS_1$ contributed by V-$3d_{z^2,x^2-y^2,xy}$ orbitals exhibits the highest $\Delta_{nk}$ values around the $H$ and $M$ points, indicated by the red color. The $FS_2$, mainly contributed by V-$3d_{xz,yz}$ orbitals, shows moderate values. Both $FS_1$ and $FS_2$ display pronounced anisotropy in the distribution of the superconducting gap. In contrast, the cylindrical $FS_3$, mainly associated with the Sb-$5p_z$ orbital, exhibits comparatively smaller and isotropic gap values, indicated by blue and cyan colors. All the Fermi surfaces are fully gapped with no nodes present. The superconducting gap distribution is presented in Fig. 4(c). The $FS_1$ and $FS_2$ develop superconducting gaps ranging from 0.3 to 0.9 meV, with an average value of 0.64 meV, while the $FS_3$ exhibits smaller gap values, centered near 0.25 meV. This distinct gap distribution provides strong evidence for orbital-selective superconducting pairing.

Moreover, by solving the superconducting gap equation (Eq. (8)) across multiple temperatures, we determined a $T_C$ of 4.3 K (Fig. 4(d)), which agrees reasonably with the experimental values of 4.2-6.6 K [8,15,26,43].

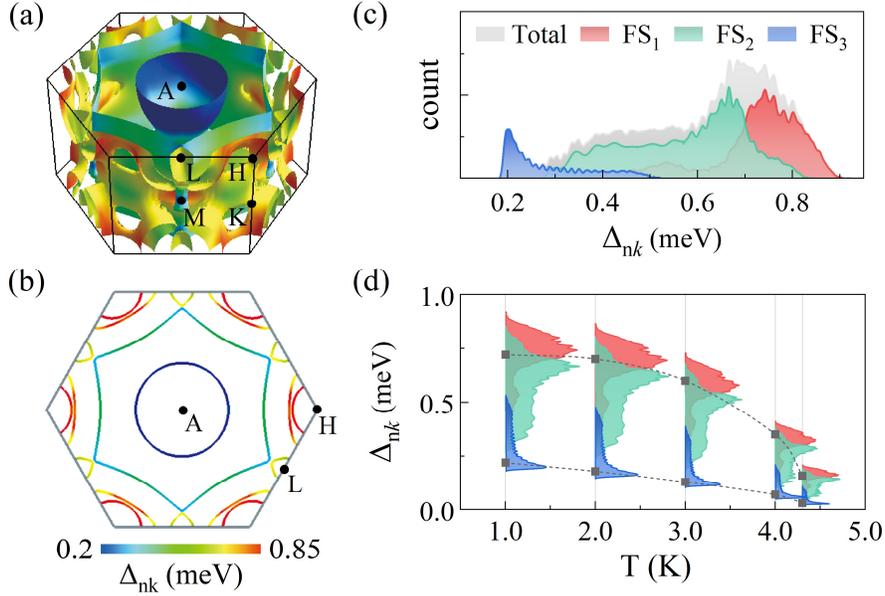

FIG. 4. Calculated superconducting gap $\Delta_{nk}$. (a) Momentum-resolved superconducting gap $\Delta_{nk}$ calculated by SCDFT. (b) Two-dimensional section of $\Delta_{nk}$ at $k_z=0.5$ plane. (c) Superconducting gap distribution on different Fermi surface sheets. (d) Temperature-dependent superconducting gap $\Delta_{nk}$. The dashed lines are guides to the eye.

As a multi-band superconductor, Fermi surfaces derived from different orbitals may give rise to multiple superconducting gaps and gap anisotropy. Such features have been experimentally observed in superconductors like FeSe [59], MgB$_2$ [60,61], and 2$H$-NbS$_2$ [62,63]. This multi-gap character of CsV$_3$Sb$_5$ has also been established experimentally [19,20,23,26,42,43]. Magnetic penetration depth [25] and electronic specific heat [45] measurements further demonstrate the coexistence of anisotropic and isotropic gaps. Our findings establish a microscopic origin for the orbital-selective two-gap superconductivity in CsV$_3$Sb$_5$. Specifically, while the V-3$d$ and Sb-5$p_z$ orbitals dominate the Fermi surface, they exhibit contrasting gap characteristics. The V-3$d$ orbitals generate a large and highly anisotropic gap, whereas the Sb-5$p_z$ orbital gives rise to a small and isotropic gap. This orbital-selective pairing mechanism not only reconciles complex experimental observations, but may also elucidate superconductivity in other kagome metals exhibiting similar multi-orbital physics.

## IV. CONCLUSION

In conclusion, our first-principles calculations incorporating SCDFT establish orbital-selective EPC as the fundamental mechanism governing the two-gap superconductivity in $CsV_3Sb_5$. The results reveal that the V-$3d$ and Sb-$5p_z$ orbitals couple selectively to distinct phonon modes, generating orbital-dependent EPC strengths. The momentum-resolved superconducting gap exhibits pronounced anisotropy and a two-gap feature. The magnitudes and anisotropic behaviors of the gaps directly correlated with orbital character. The V-$3d_{z^2, x^2-y^2, xy}$ and V-$3d_{xz,yz}$ orbitals give rise to a large, highly anisotropic gap with an average value of 0.64 meV, while the Sb-$5p_z$ orbital produces a small, isotropic gap of approximately 0.25 meV. These findings offer a microscopic understanding that unifies the experimentally observed features in $CsV_3Sb_5$ and highlighting the essential role of multi-orbital physics in governing the superconductivity in kagome metals.

## ACKNOWLEDGMENT

This work was supported by the National Key Research and Development Program of China under Contract No. 2022YFA1403203. The calculations were performed at Hefei Advanced Computing Center.

SUPPLEMENTAL MATERIAL

# Orbital-selective two-gap superconductivity in kagome metal $CsV_3Sb_5$

Run Lv[1,2], Wenqian Tu[1,2], Dingfu Shao[1], Yuping Sun[3,1,4], and Wenjian Lu[1,*]

[1]Key Laboratory of Materials Physics, Institute of Solid State Physics, HFIPS, Chinese Academy of Sciences, Hefei 230031, China

[2]University of Science and Technology of China, Hefei 230026, China

[3]High Magnetic Field Laboratory, HFIPS, Chinese Academy of Sciences, Hefei 230031, China

[4]Collaborative Innovation Center of Microstructures, Nanjing University, Nanjing 210093, China

This Supplemental Material includes:

(1) Orbital-resolved band structure and Fermi surface. (Fig. S1)

(2) Atomic-vibration-resolved phonon spectra. (Fig. S2)

*Corresponding author: wjlu@issp.ac.cn

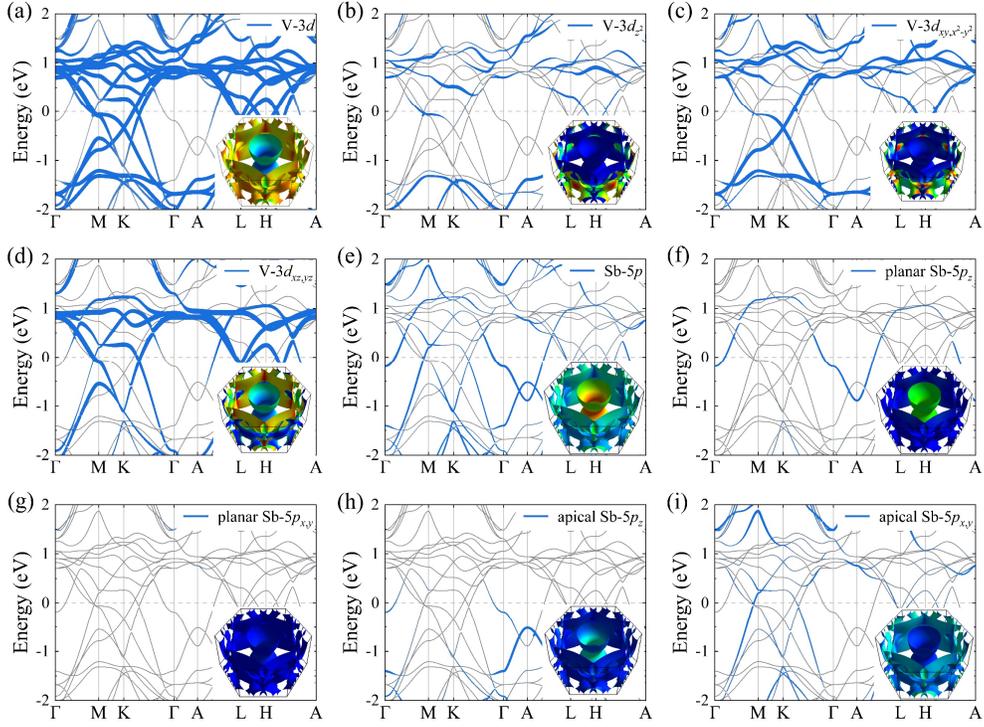

FIG. S1 Band structures and Fermi surface projected to V-3*d* and Sb-5*p* orbitals.

Comparing Figs. S1(b) and (c), one can find that the V-3$d_{xy}$ and V-3$d_{x^2-y^2}$ orbitals are hybridized with the V-3$d_{z^2}$ orbital, distributed on the FS$_1$. The FS$_2$ is dominated by the V-3$d_{xz,yz}$ orbitals (Fig. S1(d)), with minor contribution from apical Sb-5$p_{x,y}$ orbitals (Fig. S1(i)). The FS$_3$ have contribution from both planar and apical Sb-5$p_z$ orbitals (Figs. S1(f) and (h)).

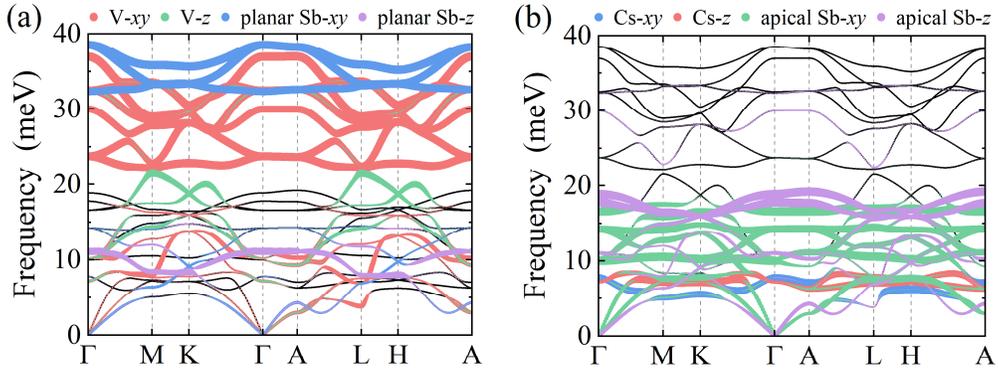

FIG. S2 Phonon dispersion spectra projected to vibration modes of different atoms.

Figure S2 presents the vibration-resolved phonon dispersion spectra of CsV$_3$Sb$_5$. The soft phonon modes at the *M* and *L* points exhibit the largest EPC strengths and are predominantly contributed by V-*xy* vibrations. The low-frequency soft phonon modes around the *L* point also receive minimal but non-negligible contribution from the out-of-plane vibrations of both Cs and apical Sb atoms, as shown in Fig. S2(b).